\documentclass[submission,copyright,creativecommons]{eptcs}
\providecommand{\event}{FROM 2024} % Name of the event you are submitting to

\usepackage{iftex}
\usepackage{array}
\usepackage{todonotes}
\usepackage{alltt}
\usepackage{listings}

\usepackage[normalem]{ulem}
\graphicspath{/Images/}
\ifpdf
  \usepackage{underscore}         % Only needed if you use pdflatex.
  \usepackage[T1]{fontenc}        % Recommended with pdflatex
\else
  \usepackage{breakurl}           % Not needed if you use pdflatex only.
\fi
\usepackage{amsmath,amssymb}
\usepackage{algpseudocode}
\usepackage{pifont}

\newcommand{\cmark}{\ding{51}}%
\newcommand{\xmark}{\ding{55}}%
\usepackage{listings, xcolor}

\definecolor{verylightgray}{rgb}{.97,.97,.97}

\lstdefinelanguage{Solidity}{
	keywords=[1]{anonymous, assembly, assert, balance, break, call, callcode, case, catch, class, constant, continue, constructor, contract, debugger, default, delegatecall, delete, do, else, emit, event, experimental, export, external, false, finally, for, function, gas, if, implements, import, in, indexed, instanceof, interface, internal, is, length, library, log0, log1, log2, log3, log4, memory, modifier, new, payable, pragma, private, protected, public, pure, push, require, return, returns, revert, selfdestruct, send, solidity, storage, struct, suicide, super, switch, then, this, throw, transfer, true, try, typeof, using, value, view, while, with, addmod, ecrecover, keccak256, mulmod, ripemd160, sha256, sha3}, % generic keywords including crypto operations
	keywordstyle=[1]\color{blue}\bfseries,
	keywords=[2]{address, bool, byte, bytes, bytes1, bytes2, bytes3, bytes4, bytes5, bytes6, bytes7, bytes8, bytes9, bytes10, bytes11, bytes12, bytes13, bytes14, bytes15, bytes16, bytes17, bytes18, bytes19, bytes20, bytes21, bytes22, bytes23, bytes24, bytes25, bytes26, bytes27, bytes28, bytes29, bytes30, bytes31, bytes32, enum, int, int8, int16, int24, int32, int40, int48, int56, int64, int72, int80, int88, int96, int104, int112, int120, int128, int136, int144, int152, int160, int168, int176, int184, int192, int200, int208, int216, int224, int232, int240, int248, int256, mapping, string, uint, uint8, uint16, uint24, uint32, uint40, uint48, uint56, uint64, uint72, uint80, uint88, uint96, uint104, uint112, uint120, uint128, uint136, uint144, uint152, uint160, uint168, uint176, uint184, uint192, uint200, uint208, uint216, uint224, uint232, uint240, uint248, uint256, var, void, ether, finney, szabo, wei, days, hours, minutes, seconds, weeks, years},	% types; money and time units
	keywordstyle=[2]\color{teal}\bfseries,
	keywords=[3]{block, blockhash, coinbase, difficulty, gaslimit, number, timestamp, msg, data, gas, sender, sig, value, now, tx, gasprice, origin},	% environment variables
	keywordstyle=[3]\color{violet}\bfseries,
	identifierstyle=\color{black},
	sensitive=true,
	comment=[l]{//},
	morecomment=[s]{/*}{*/},
	commentstyle=\color{gray}\ttfamily,
	stringstyle=\color{red}\ttfamily,
	morestring=[b]',
	morestring=[b]"
}

\newcommand{\ProgramVariableSet}{Var}
\newcommand{\IntervalsSet}{Int}
\newcommand{\IntervalsSetExtended}{IntExt}
\newcommand{\AnalysisInformation}{AI}
\newcommand{\PaperTitle}{Leveraging Slither and Interval Analysis to build a Static Analysis Tool}

\title{\PaperTitle}
\author{Ștefan-Claudiu Susan
\institute{Alexandru Ioan Cuza University of Iași,\\ Department of Computer Science\\
Iași, România}
\email{claudiu\_susan@yahoo.com}
}
\def\titlerunning{\PaperTitle}
\def\authorrunning{Ștefan-Claudiu Susan}
\begin{document}
\maketitle

\begin{abstract}
Even though much progress has been made in identifying and mitigating smart contract vulnerabilities, we often hear about coding or design issues leading to great financial losses. This paper presents our progress toward finding defects that are sometimes not detected or completely detected by state-of-the-art analysis tools. Although it is still in its incipient phase, we developed a working solution built on top of Slither that uses interval analysis to evaluate the contract state during the execution of each instruction. To improve the accuracy of our results, we extend interval analysis by also considering the constraints imposed by specific instructions. We present the current solution architecture in detail and show how it could be extended to other static analysis techniques, including how it can be integrated with other third-party tools. Our current benchmarks contain examples of smart contracts that highlight the potential of this approach to detect certain code defects. 
\end{abstract}

\section{Introduction}
In the aftermath of the 2008 financial crisis, the population started losing faith in banks, financial governing authorities, and fiat currencies. This was the perfect context for introducing Bitcoin\cite{Bitcoin}, the first cryptocurrency and the first practical implementation and use case for Blockchain technology. It was in complete opposition with the traditional financial instruments, promising transparency, decentralization, and immutability.

Unlike Bitcoin, which was a purely financial implementation of blockchain, Ethereum \cite{ethereum} also supported deploying programs on its network, making it a Blockchain Software Platform. These programs are known as smart contracts. As the name suggests, smart contracts contain a digital form of an agreement made between two or more parties. In contrast to traditional contracts, due to their automatic enforcement of terms, they do not require any trust between the involved parties. Along with the automation capabilities specific to conventional software products, the immutability, transparency, and decentralization of the blockchain make smart contracts the perfect fit for such a use case. The most popular programming language for implementing such contracts is Solidity ~\footnote{Solidity, version 0.8.26: \url{https://docs.soliditylang.org/en/v0.8.26/}}.\\

As with any incipient technology, Solidity and smart contracts in general had many faults when first introduced. These defects were quickly exploited by malicious users. These issues mainly came from the language design. Two such example are ``The DAO Hack``\cite{DAO} and ``Parity Wallet Multisig Bug``\cite{ParityWallet}
Incidents such as these motivated researchers and other people in the community to start discovering and classifying the types of vulnerabilities that can appear in smart contracts. These defects include the ones common among ``classic`` programming languages such as \texttt{Division by zero} or \texttt{Integer Underflow/Overflow}, as well as bugs that are specific to the Solidity language and the Blockchain environment, such as \texttt{Reentrancy} and \texttt{Transaction State Dependency}. Taxonomies such as the one presented in \cite{rameder2022review} were created with the purpose of offering a better understanding of what bugs might appear. Community-maintained taxonomies such as SWC Registry\footnote{SWC Registry:\url{https://swcregistry.io/}} and DASP Top 10\footnote{DAPS Top 10 \url{https://dasp.co/}} must also be mentioned as notable efforts in this area.

In addition to classifying the most common issues, a lot of effort was also put into starting to manually audit smart contracts and developing tools capable of automatically detecting bugs. These tools implement a variety of approaches to signal code defects, such as symbolic execution and static program analysis. Such implementations include Slither\cite{slither}, a tool that implements a variety of detectors, Securify\cite{Securify}, a static analysis tool that models good and bad coding patterns as logic formulas; Solhint\cite{Solhint}, a linter for the Solidity language; Mythril\cite{Mythril}, a symbolic execution approach that analyzes EVM (Ethereum Virtual Machine) Bytecode. Even though not an analysis tool in itself, we must also mention the static analysis plugin that Remix \footnote{Remix IDE: \url{https://remix-project.org/}}, the most popular IDE for Ethereum development, offers. During a previous study we conducted\cite{taxanomyPaper}, Slither emerged as the overall best-performing tool. Even though it had the highest overall score, there were still a considerable amount of vulnerabilities that were not detected by it or any other tool included in our study. This made us analyze the way it is implemented in greater detail and research methods to aid it in providing more robust detections.\\

In this paper, we present our current progress in implementing a new analysis tool that is built on top of Slither and leverages the information provided by it after parsing the contract. Our current approach is based on \emph{interval analysis}, with the possibility of being extended in the future. Using this technique, we can approximate the range of values of each variable during each program point, regardless of whether it is a state variable, parameter, local variable, or built-in variable. Having these approximations, we are then able to detect possible coding or design defects based on program state and constraints. A good example of such issues is an unreachable program statement; if a \texttt{require} statement contains a condition that can never be fulfilled, then all the instructions after that will not be executed. This also applies to unreachable if branches or loop bodies.
Even though our tool is able to approximate the program state for a Solidity function with a reasonable degree of accuracy, it has certain limitations. One such limitation lies in the fact it is not yet able to interpret every instruction in Solidity. Currently, our tool only supports intraprocedural analysis. Another, more specific to interval analysis, is the limited accuracy of the approximation without user input.
\paragraph{Summary of contributions}
\begin{enumerate}
\setlength\itemsep{0em}
    \item We provide an in-depth explanation of how Slither can be extended and its main modules.
    \item We present the architecture that we implemented and the means used to connect with external modules.
    \item We provide a collection of smart contracts that our tool can run against and evaluate our solution.\\[-5ex]
\end{enumerate}
\paragraph{Paper organisation}
Section~\ref{sec:relatedwork} contains a summary of our previous work in this area as well as a short presentation of other state-of-the-art analysis tools. Section~\ref{sec:background} a briefing of the Static Program Analysis theories that serve as the building blocks of our work. In Section~\ref{sec:Slither}, we describe how Slither can be used in a custom implementation and describe the main modules and data types included. We present our solution architecture in Section~\ref{sec:solutionarchitecture}. We evaluate our solution in Section~\ref{sec:evaluation} and present its limitations in Section~\ref{sec:limitations}. The paper concludes with Section~\ref{sec:conclusion}.

\section{Related work}
\label{sec:relatedwork}
We already presented the fundamentals of our approach in our initial progress report~\cite{From2023Paper}. It serves as the building blocks of our current implementation. Even though we are still facing some of the limitations outlined during our previous report, we managed to overcome a few important limitations, such as handling conditional, repetitive, and the most widely used builtin statements. In addition, our current implementation is considerably more robust and easy to extend.

There are a number of remarkable tools, besides Slither, that perform static analysis while implementing different approaches, among them, we must mention:
\begin{enumerate}
    \item Securify\cite{Securify} is a smart contract security scanner that heavily relies on First Order Logic concepts and formulas to detect issues. It defines multiple patterns as logical formulas and checks program statements to see if any of them match. There are compliance and violation patterns. If a program section matches a compliance pattern, it is not flagged as a defect. If it matches a violation pattern, it is marked as an error with a high degree of certainty. If none of the previous situations apply, it is marked as a warning.
    \item Solhint\cite{Solhint} is an open-source linter for the Solidity language. It can be installed via \texttt{npm} while also being available as a plugin for numerous IDEs. Its behaviour relies on detection rules set by the user. These guidelines include \emph{Security Rules}, \emph{Style Guide Rules} and \emph{Best Practices Rules}. It requires a configuration file that declares which detection rules must be employed. A default configuration can be generated using a \texttt{npm} command.
    \item Remix\cite{Remix} is a dedicated IDE for developing DApps (Decentralized applications). It provides many useful features that can aid in implementing such applications. Besides the compiler whose version can be selected by the user, it also features an emulated version of the Ethereum network. 
    By default, there are 10 available addresses, each holding 100 Ether, any of these addresses can be used to deploy the contracts. After deployment, any address can be used to call functions located in those contracts. Besides integration with Solhint and Slither, it also features its own static analysis plugin. We were unable to find any details regarding the techniques employed in its implementations. During a previous benchmark that we conducted on a number of analysis tools, we found that the static analysis plugin for Remix yielded notable results compared to most tools. 
\end{enumerate}

Although it relies more on symbolic and concrete execution, Hevm\cite{Hevm} also aims to detect reachable and unreachable final states for a smart contract function. This solution determines the final states from symbolic inputs. The final states are statically determined to be unreachable or marked as potentially reachable. For the former, SMT queries are generated and passed to a solver. An approach similar in the use of a solver to check the satisfiability of states using contracts is also featured in our solution. Hevm can also check the equivalence of two different implementations for the same functionality by checking the equivalence of final states.

We must also mention Simbolik\footnote{Simbolik is available at \url{https://simbolik.runtimeverification.com/}}. Even though it is a debugging tool that uses symbolic execution, not a static analysis code defect detection tool, it needs to be mentioned due to its capabilities of approximating the program state for any step. Leveraging the capabilities of K Framework\cite{KFramework}, it can be used for debugging, symbolic testing, and detecting path conditions.

\section{Background}
\label{sec:background}
\subsection{Dataflow Analysis}
Dataflow Analysis\cite{allen1976program} is a program analysis paradigm that was developed in the context of optimizations performed by compilers and it remains one of its most prominent use cases~\cite{DataflowBook}. Being a static analysis technique, it aims to simulate the behaviour of a program at runtime without executing it. As the name suggests, it analyzes the flow of information through the control flow graph of a program by collecting data about the possible values and states of variables and expressions at various points of a program.

We should clearly differentiate it from \emph{control-flow-analysis}\cite{ControlFlowAnalysis}. It clearly determines the order of instructions that are executed in an algorithm given a certain context. They are different in their purpose, but they rely on each other's findings. To properly determine the flow of data, the order of operations is needed as it influences the concrete values at concrete program locations. On the other hand, to determine the order of operations when runtime decisions are present, an approximation of data state is necessary. 

Based on different characteristics, data-flow analysis techniques can be divided into different categories. Some of the classification criteria are the following\cite{PrincipelsOfProgramAnalysis}:

\begin{itemize}
    \item Depending on the direction of flow: \textbf{top-down (forward) analysis} that follows the actual execution flow of the program, or \textbf{bottom-up (backward) analysis} that does the opposite,
    \item Depending on the scope: \textbf{intraprocedural analysis} that only considers a single function and \textbf{interprocedural analysis} that takes into consideration the interaction between functions,
    \item Depending on program flow consideration: \textbf{flow-sensitive analysis} that considers program flow and \textbf{flow-insensitive analysis} that does not take it into consideration,
    \item Depending on how execution paths intersection is treated: \textbf{may analysis} that performs program state union and \textbf{must analysis} that performs intersection.
\end{itemize}

Examples of data-flow analysis techniques include, but are not limited to, \emph{Live Variables}, \emph{Available Expressions}, \emph{Constant Propagation}, etc. described in more detail in~\cite{DataflowSystemNotes}. A data-flow analysis instance can be formalized as a system containing the following elements:

Given a program $P$, let $G(P)=(V,E,\mathit{cmd})$ be the control flow graph of P where:
\begin{itemize}
    \item $V$ is the set of vertex from the graph,
    \item $E \subseteq V \times V$ is the set of edges from the graph,
    \item $\mathit{cmd}$ is the set of statements from P.
\end{itemize}

A data-flow system $S =(\mathit{Lab}, \mathit{Ext}, \mathit{Flw}, (D,\sqsubseteq), \epsilon, \varphi)$
\begin{itemize}
    \item A set of program labels $\mathit{Lab}=V$ in $G(P)$. For example, the numerical labels attributed to each vertex of the control flow graph,
    \item A subset of extremal labels, $\mathit{Ext}$, the labels where the program starts or where it ends depending on the flow. The nodes marked with extreme labels either have only outbound or only inbound edges,
    \item A flow relation between labels, a function that models the transition between the program points. $\mathit{Flw}$: $\mathit{Flw}= \mathit{E}$ for forward analysis and $\mathit{Flw}= \mathit{E}^{-1}$ for backward analysis,
    \item A complete lattice\cite{completeLattice} $(D,\sqsubseteq)$ containing all the possible state values, this will serve as the domain of analysis information values. It must contain $\bot$ (bottom) and $\top$ (top) values. These elements are the greatest-lower and least-upper bounds of $D$,
    \item An extremal value $\epsilon\in D$ for extremal labels,
    \item A collection of transfer functions $\varphi_\ell:D\to D$, $\ell\in\mathit{Lab}$. These functions reflect the changes in a program's state produced by a statement.
\end{itemize}
 
 Even though the analysis information is not the same for all analysis types, it can be represented as key-value pairs, where the key is the label of the program point and the value is the state representation relevant to that type of analysis. In other words, the analysis information for the whole program contains a collection of representations for each statement/node, which in turn contains a collection of representations for each variable.

This type of static analysis works by presenting the program state as a system of equations and trying to calculate the least fixed points\cite{mccarthy1959basis}. To accomplish this, several algorithms have been developed, including the MOP ( Meet Over all Paths)~\cite{kildall1973unified} and the Worklist~\cite{PrincipelsOfProgramAnalysis}.
We will focus on the latter since it is also the one that our solution implements. 

It works by iterating over the control flow graph until no changes occur in the program state. The worklist is initialized with the extremal labels and checks adjacent nodes according to the flow function. Initially, the algorithm could run for an infinite number of iterations for some programs, especially the ones containing loops. This happens when the complete lattice in the data-flow system does not satisfy ACC (Ascending Chain Condition).

We consider that a complete lattice $(D,\sqsubseteq)$ satisfies ACC if the sequence $d_0 \sqsubseteq d_1 \sqsubseteq d_2 \sqsubseteq \cdots \subseteq d_n$ eventually stabilizes. The chain stabilises if $ d_n = d_{n+1}=\cdots$ for some $n\ge 0$. Knowing this, we can certainly say that the algorithm will complete after $|Lab| * n$ iteration at most. Non satisfying ACC domains do not grant this certainty and the algorithm might run indefinitely.

This issue was addressed by using widening operators\cite{cortesi2011widening}. Even though they provided an over-approximation, their use made sure that the algorithm runs over a finite number of iterations. To obtain a more precise solution, narrowing operators have been employed. Using them, a more precise approximation can be obtained. 

\subsection{Interval Analysis}
\label{subsec:IntervalAnalysis}
Interval analysis\cite{IntervalAnalysis1979}\cite{IntervalAnalysis} is a technique of static analysis that aims to approximate the value of each program variable using intervals. The classic approach to interval analysis uses numeric intervals to represent the range of values of variables at each point in the program.

It is an instance of abstract interpretation\cite{abstractinterpretation77}\cite{abstractinterpretation}. It abstracts precise numerical values into intervals. With a simplified domain and operations, this technique can be used to effectively reason a program's potential behavior without necessitating concrete execution and exact values.

A data-flow system $S =(\mathit{Lab}, \mathit{Ext}, \mathit{Flw}, (D,\sqsubseteq), \epsilon, \varphi)$ for interval analysis using numeric intervals, where \ProgramVariableSet is the set of variables from the program and $Int$ is the set of numeric intervals, is the following:

\begin{itemize}
    \item $Lab: Lab$=V in G(P),
    \item $\mathit{Ext}$: $\mathit{Ext}$=$\{V \mid \nexists V_i, V_i \rightarrow V \in Edges \}$ (program starting point, forward analysis),
    \item $\mathit{Flw}$: $\mathit{Flw}= \mathit{Edges}$ (forward analysis),
    \item $(D,\sqsubseteq)$: D=$\{\AnalysisInformation \mid \AnalysisInformation:\ProgramVariableSet \rightarrow \IntervalsSet \}$, $\AnalysisInformation_1 \sqsubseteq \AnalysisInformation_2$ iff $\AnalysisInformation_1(x) \subseteq \AnalysisInformation_2(x)$ for every $x \in \ProgramVariableSet$. The analysis information for a program point contains the variables that are accessible from the scope of that statement. It maps each variable to a numeric interval. This domain only contains all possible values for \texttt{int} and \texttt{bool}. To be able to handle a program with more complex types, this domain must be extended,
    \item $\epsilon = [-\infty,\infty]$. This is the most general approximation for variables that hold numerical values. In practice, depending on the data type, this interval can be narrowed down. For example, $\epsilon = [0,\infty]$ for unsigned data types and $\epsilon = [0(false),1(true)]$ for boolean values,
    \item the transfer function:
\[
\varphi_\ell = 
\begin{cases} 
    \AnalysisInformation & \text{if $\ell$ does not modify state} \\
    \AnalysisInformation[x_i = [v,v]] & \text{if $\ell$ modifies the variable $x_i$ with the value $v$}
\end{cases}
\]
\end{itemize}

To further improve the accuracy of our approximations, we can make use of constraints imposed by certain statements. Such instructions are common among most programming languages, such as the \texttt{if} statement through its conditional branches and loops via their invariant condition. We must also mention that Solidity has two built-in instructions that impose constraints: \texttt{require} and \texttt{assert}. 

 In the domain defined below, $Val$ holds the variables state and $Con$ is the set of constraints for that point in the program. 

 To be able to handle complex types, we must extend the codomain of $Val$, we will note it as $\IntervalsSetExtended$. In addition to numeric intervals, we must also add mappings between collection indexes or struct fields and numerical intervals. This means that we need to define $\IntervalsSetExtended$ recursively.
 
Considering a variable $v$ that has a data type $Type$, we define $Val(v)$ as follows:\\
The possible values for $Type$ are a subset of the data types included in Solidity~\footnote{Solidity 0.8.26 data types: \url{https://docs.soliditylang.org/en/v0.8.26/types.html}}..\\
 $Type = int \mid bool \mid Struct[ f1 : Type_1, . . . , fn : Type_n] \mid Array[Type] \mid Map[Type_1, Type_2]$.
\begin{itemize}
    \item $Val(\texttt{int}) = \IntervalsSet$,
    \item $Val(\texttt{bool}) = \{[0,0],[1,1],[01]\}$,
    \item $Val(\{f_1:T_1,\ldots, f_n:T_n\}) = \{\{\{f_1:v_1,\ldots, f_n:v_n\}\mid v_i\in Val(T_i), i=\overline{1,n} \}$ where $f_1,...,f_n$ are all the fields defined in the structure,
    \item $Val(Array[T]) =\{(i_1:v_1,\ldots,i_n:v_n)\mid  i_j\in Val(\texttt{int}), v_j\in Val(T), j=\overline{1,n}, n\ge 1\}$ and n is the length of the array. If the array has been declared without a size and has not been initialized, we consider n=0,
    \item $Val(Map[T_1,T_2]) =\{(k_1:v_1,\ldots,k_n:v_n)\mid  k_j\in Val(T_1), v_j\in Val(T_2), j=\overline{1,n}, n\ge 1\}$ where $k_1,..,k_n$ are the keys that have been assigned a value up to that point of the program.
\end{itemize}
 For reference types that contain fields of various data types, Val is recursively applied to each field according to it's data type. We must also extend $\subseteq$(inclusion), $\cup$(reunion) and $\cap$(intersection) for our new domain. For complex types, these operators will be applied in two steps. First, on the keys, and then on the underlying values. \\
 We will consider $K$ as the set of keys for a complex data structure. For example, $K(s)$ if s has the data type of \texttt{Struct} is the set of all fields declared in that struct. For mapping and arrays, the result of $K$ is the set of keys/indexes that have been initialized up to that point. In addition, for arrays of fixed length, the result of $K$ will \\
 The behavior for these extended operations is the following:

 \begin{itemize}
          \item $Val(x_1)\subseteq Val(x_2)$ if $K(x_1) \subseteq K(x_2)$ and $Val(x_1)[k_i] \subseteq Val(x_2)[k_i] \forall k_i\in K(x_1)$
          \item $Val(x_1) \cap Val(x_2) = \{k_i:Val(x_1)[k_i] \cap Val(x_2)[k_i]\} \forall k_i \in K(x_1) \cap K(x_2)$
           \item $Val(x_1) \cup Val(x_2) = \{k_i:Val(x_1)[k_i] \cup Val(x_2)[k_i]\} \forall k_i \in K(x_1) \cup K(x_2)$. If $k_i$ is not included in $K(x_1)$ and is included in $K(x_2)$ then only the value $Val(x_2)[k_i]$ will be considered when computing the result and vice-versa. 
 \end{itemize}

Considering $BExp$ as the set of boolean expressions present in the program, our domain becomes the following:
\begin{center}
 D=$\{\AnalysisInformation=(Val,Con) \mid Val:\ProgramVariableSet \rightarrow \IntervalsSetExtended, C \subseteq BExp\}$,
 
 $\AnalysisInformation_1 \sqsubseteq \AnalysisInformation_2$ iff $Val_1(x) \subseteq Val_2(x)$ for every $x \in \ProgramVariableSet,   Con_1 \implies Con_2$.

 $\AnalysisInformation_1 \cup \AnalysisInformation_2=(Val_1 \cup Val_2, Con_1 \cup Con_2)$
 \end{center}
 
The constraints are defined over symbolic variables. They are propagated only downward when traversing the CFG. Currently, Solidity does not feature a \texttt{Goto} instruction or any other jump statement. Due to this, the only possible cycles that can be present in the CFG are the ones created by loops. To mediate cases where a node can be reached from multiple paths, we only impose constraints on the starting point of the loop that come from the initial path of execution, we do not add the constraints imposed during the loop body or the loop invariant. We consider that this approach is the most suitable for our use case since it ensures termination when the program includes loops. Although it might lead to a loss of precision because fewer constraints are added, it is the most stable approach.  

If we take this optimization into consideration, the $\varphi$ function becomes the following:

We consider the analysis information for the current node as $\AnalysisInformation=(Val,Con)$. The 
\[
\varphi_\ell = 
\begin{cases} 
    (Val,Con) & \text{if $\ell$ does not modify state and does not impose or remove constraints} \\
    (Val[x_i = [v,v]],Con) & \text{if $\ell$ modifies the variable $x_i$ with the value $v$} \\
    (Val,Con\cup C_1)  & \text{if $\ell$ imposes the constraint $C_1$ and Lab(source) < Lab(destination) }  \\ 
    (Val,Con\setminus C_1) & \text{if $C_1$ was enforced and $\ell$ invalidates it or if the scope of $C_1$ ends}

\end{cases}
\]
The constraints are required to check whether the current state is satisfiable. We make use of an SMT solver to perform this check. Our solution is designed to support a variety of solvers in a plugin approach. At the time of writing, only integration with Z3 was implemented. This interaction is explained in further detail in the following sections of this paper.

To highlight how this analysis technique approximates the state of a program, we provide the following example:
\begin{lstlisting}[language=Solidity]
    function magicNumber(uint x) pure external returns(uint){
       uint index=0; //statement 1
       uint value=x; //statement 2
       require(x<15); //statement 3
       while(index<x) //statement 4
       {
            if(index\%2==0) //statement 5
            {
                value=value*2; //statement 6
            }
            else
            {
                value=value*3; //statement 7
            }
             x=x+1; //statement 8
       }
       return value; //statement 9
    }
\end{lstlisting}

The approximations of state for each program point are displayed in Table~\ref{tab:exampleProgramTable}.
\begin{table}[h]
    \centering
    \begin{tabular}{|c|c|c|c|c|}
        \hline
        \textit{Statements} & \textit{x} & \textit{index} & \textit{value} & \textit{Constraints} \\ \hline \hline
        1 & [0,$\infty$] & $\emptyset$ & $\emptyset$ & \{\} \\ \hline
        2 & [0,$\infty$] & [0,0] & $\emptyset$ & \{\}\\ \hline
        3 & [0,$\infty$] & [0,0] & [0,$\infty$] & \{\} \\ \hline
        4 & [0,$\infty$] & [0,$\infty$] & [0,$\infty$] & \{x<15\} \\ \hline
        5 & [0,$\infty$] & [0,$\infty$] & [0,$\infty$] & \{x<15 \&\& index<x\} \\ \hline
        6 & [0,$\infty$] & [0,$\infty$] & [0,$\infty$] & \{x<15 \&\& index<x \&\& index\%2==0\} \\ \hline
        7 & [0,$\infty$] & [0,$\infty$] & [0,$\infty$] & \{x<15 \&\& index<x \&\& index\%2!=0\} \\ \hline
        8 & [0,$\infty$] & [0,$\infty$] & [0,$\infty$] & \{x<15\} \\ \hline
        9 & [0,$\infty$] & [0,$\infty$] & [0,$\infty$] & \{x<15\} \\ \hline
        End & Data 9.2 & [0,$\infty$] & [0,$\infty$] & \{x<15\} \\ \hline
    \end{tabular}
    \caption{Interval analysis with constraints for the \texttt{magicNumber} function.}
    \label{tab:exampleProgramTable}
\end{table}

\section{Slither}
\label{sec:Slither}
\subsection{Overview}

Slither\cite{slither} is a static analysis tool that offers security and good practices advice for smart contracts. Compared to other analysis tools, it runs remarkably fast. The tool obtains information by using the Solidity compiler and converting EVM bytecode into a proprietary intermediate representation called SlithIR. This new representation can be visualized using a printer, it also has an SSA (Static Single Assignment) variant. Before converting the initial code to SlithIR, it uses the information provided by the compiler to construct the CFG, Inheritance Graph. This is achieved by parsing the AST (Abstract Syntax Tree) resulting from contract compilation.

At the time of writing, Slither features 93 detectors\footnote{A detailed list is available on the Slither GitHub repository:\url{https://github.com/crytic/slither}}. These detectors are self-contained plugins that process the information compiled independently. Being an open-source tool, any user can contribute with new detectors. To do so, the contributor needs to extend an abstract class that defines the common behaviour among all detectors: processing contract information (i.e., detecting the bug); displaying the result; and attributes regarding the detector documentation (severity, confidence, etc).

In addition to detection purposes, the contract information can also be displayed using `printers`.
They display contract information internally gathered by Slither in a human-readable format. In the official documentation, the printers are divided into two categories: `Quick Review Printers` and `In-Depth Review Printers`. The printers that we consider offer the most useful insights are the following: human-summary; contract-summary; loc (lines of code); cfg (control-flow-graph) and function summary.

Besides detecting issues and printing contract information, Slither also has many other capabilities. These features include: generating reports; generating code, extracting an interface from an existing contract; correcting code, either reformatting the contract, fixing simple vulnerabilities automatically or flattening the codebase; code checks, and checking compatibility with well-known standards or with newer features of Solidity.

The tool needs to be installed as a Python module via pip\footnote{\url{https://pypi.org/project/slither-analyzer/}}. After the installation is performed, the user has access to the command line tool and the Python modules that can be included in a custom implementation. The GitHub page features a generous developer documentation\footnote{Slither API documentation:\url{https://crytic.github.io/slither/slither.html}}. The main entry point for a new implementation is a class called `Slither`. An object of this type needs to be initialized by providing the path to the Solidity files that contain the targeted contract(s). After obtaining an instance of that class, the developer has access to many useful details that can be used to implement a new analysis tool or add a new detector to Slither. This information is presented as nested objects. The main object contains a list of contracts, each contract contains a list of functions, each function contains a list of nodes, and so on. Besides the children list, each object that models a contract element holds specific information, such as state variables for contracts and parameters for functions.

\subsection{Limitations}
To our best knowledge, Slither does not include the possibility of querying the contract state for each statement of a function, or at least at the start and end of execution. It does not implement interval analysis or other techniques capable of approximating the contract state. We consider that such a feature could be integrated into one of Slithers printers, aid in identifying code defects that are currently not detected, or improve the detection rate for issues that are already detected.

In addition to that, it appears that Slither does not take into consideration constraints imposed by previous statements. Constraints are added either by traditional conditional statements such as an \texttt{if} statement, or by instructions such as \texttt{assert} or \texttt{require}. Due to this, Slither is not always able to identify unreachable conditional branches. The unreachable branches could be an if/else block, the body of a loop with a contradiction acting as its condition, or even whole function sections if they are preceded by an \texttt{assert} or \texttt{require} statement that will never pass.

We present the following example:
\begin{lstlisting}[language=Solidity]
pragma solidity 0.8.23;

contract BidContract {
    mapping(uint=>uint) public  bidders;
    function bid(uint bidderNumber) public  payable {
        require(msg.value>10);
        uint newBid=bidders[bidderNumber]+msg.value;
        if(newBid>10)
        {
            //Since msg.value>10 implies that newBid>10,
            //this brach will always execute
             bidders[bidderNumber]=newBid;
        }
        else 
        {
            //Since the "then" branch is based on a tautology,
            //this branch will never execute
            revert("Inssuficient bid");
        }
  }
 }
\end{lstlisting}
\textbf{Obs:} Since we are working with unsigned integers, the variables cannot contain negative values.

If used on the contract above, Slither will signal some issues that are indeed present. The first issue relates to the fact that the version of Solidity that was used to compile it is too recent to be used. The other problem detected by Slither targets the absence of a function that allows the user to withdraw ether, we did not provide one in this contract because it is not relevant to our example. None of these defects are related to the problem that we present using this contract.

However, it will not address the fact that the \texttt{else} branch is unreachable.

Currently, our efforts are focused on this area. We aim to provide an implementation capable of approximating the program state as precisely as possible and identify issues based on the analysis information that we gather. In addition to that, we implemented a constraint system. Using a third-party solver, we can find code blocks that will never execute because they are on unreachable conditional branches.

\textbf{Observation}: We must also highlight an issue in Slithers parsing of the above contract that we discovered during our solutions development. If the \texttt{then} block of an \texttt{if} statement is empty and the \texttt{else} block is not, Slither will treat the latter as the "true" branch and the statement after the conditional statement as the "false" branch. To obtain a correct result, the truth values should be reversed. Bug encountered in \textbf{version 0.10.0} of the Slither pip package.  

\section{Solution architecture}
\label{sec:solutionarchitecture}
Our architecture can be implemented using any language that offers support for the Object Oriented programming paradigm as long as the necessary third-party integrations are available. In the following presentation and diagrams, we use Slither as the contract parser and Z3 for the constraint solving component. These are the integrations that we implemented so far, but our architecture is easily extensible to other third-party solutions that fulfill similar roles. Currently, we are able to analyze the \textbf{SSA} (Static single-assignment) version of a program, as well as its \textbf{Non-SSA} form. We found the SSA variant to be more convenient since SlitherIR also has a SSA version.

\begin{figure}
\centering
    \includegraphics[scale=0.55]{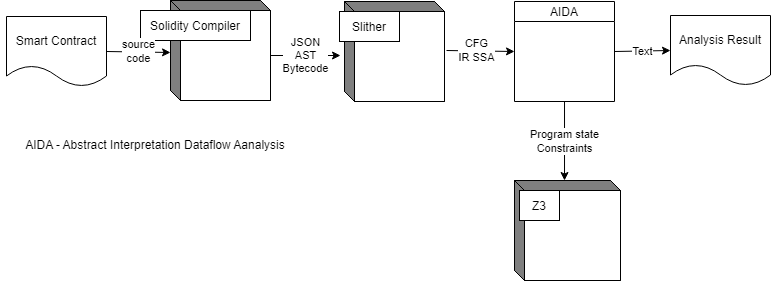} 
    \caption{System context diagram of AIDA}
    \label{fig:SolutionSystemDiagram}
\end{figure}

In Figure~\ref{fig:SolutionSystemDiagram}, we present how our system receives input data, the interaction with the components that it depends on and how it produces the output data. The flow of execution is the following:
\begin{enumerate}
    \item The user provides the path to a Solidity file along with the names of a contract and a function;
    \item Use Slither to parse the contract;
    \item Extract the information that it needs from the parsing result;
    \item Run the least fixed point algorithm. Currently, only the worklist algorithm is implemented;
    \item Translate constraints and program state into a representation compatible with the solver and call it;
    \item Write the program state for each statement of the function into a file;
    \item The user can manually analyze the program state generated by our solution.
\end{enumerate}

\begin{figure}[h]
\centering
    \includegraphics[scale=0.5]{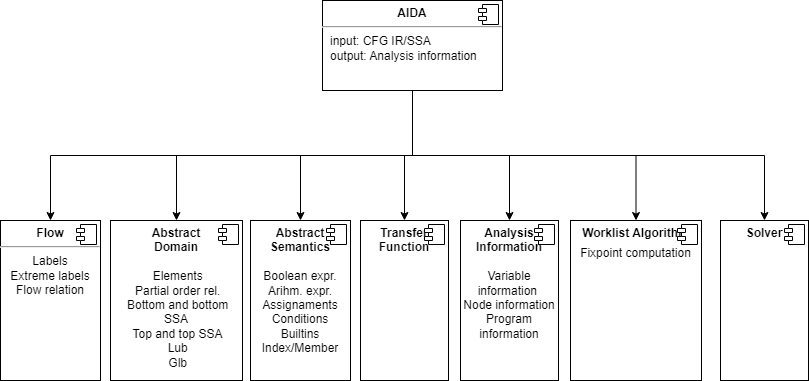}
    \caption{Modules of AIDA}
    \label{fig:SolutionArchitecture}
\end{figure}

Even though our solution currently only supports interval analysis and is integrated only with Z3 and Slither, we designed it to be extensible from the beginning. We achieved this by leveraging well-known software design principles and design patterns such as Adapter, Composite and Template Method. We make use of abstractions to define the general behavior, input and output that we expect a static analysis implementation to have. The same principle applies to external integrations, we define the data that we expect to extract from a contract parser. Figure~\ref{fig:SolutionArchitecture} provides a diagram that contains the most important modules of our implementation. All modules currently contain an abstract definition as well as concrete implementations for Slither and Z3.

The labels, extremal labels and flow relation between nodes are computed in a similar fashion to the one described in Subsection ~\ref{subsec:IntervalAnalysis}. We determine the extremal labels and flow relation via graph search algorithms that run on the CFG data structure provided by Slither. 

We defined our own data structure to store information about a program variable. 
Along with the interval, we also store other useful information such as its name, type, scope, declaring node, if it is a reference type variable or not, etc. There are some special cases where some variables might require more information to be stored than others. For this case, we also defined an "additional\_information" field as a dictionary. The volume of information in this field depends on the type of variable and its usage. For example, boolean variables that are used in constraints require more information about them to be stored for a proper interpretation of that constraint.

Since our domain does not satisfy ACC, we needed to implement widening operators. This is necessary to ensure termination when the program contains loops. Our implementation for the least upper bound for two simple numeric intervals is presented in the code snippet below:

\begin{lstlisting}[language=Python]
def get_least_upper_bound_numeric_interval(first_element: NumericIntervalApproxValue, second_element: NumericIntervalApproxValue):
    if len(first_element) == 0:
        return second_element
    if len(second_element) == 0:
        return first_element
    lower_bound = first_element[0] if first_element[0] <= second_element[0] else float(
        "-inf")
    upper_bound = first_element[1] if first_element[1] >= second_element[1] else float(
        "inf")
    return NumericIntervalApproxValue((lower_bound, upper_bound))
\end{lstlisting}

Our implementation also supports complex types, such as arrays, mappings, and structures. Unlike scalar variables where the intervals can be simply represented as a pair of two numeric values, complex types require a different type of representation. Due to this, we modeled the intervals for complex types as dictionaries. This approach offers us a great deal of flexibility, we can use numerical indexes as keys for arrays or the field names for structs for example. For mappings, this representation comes naturally.

Even though our implementation supports nested complex types, these are not seen often. Since Ethereum requires a gas fee to run smart contracts, the code must be as resource efficient as possible, this is especially true for execution time and the amount of storage required. This mechanism heavily discourages developers from implementing complex scenarios and data structures  

The way we designed our representation of program state for interval analysis is very similar to the one described in Section~\ref{sec:relatedwork}. The only difference lies in the fact that the information for a node does not contain a single collection of intervals corresponding to each variable. The information is distributed in the following categories: numeric variables, boolean variables and constraints. Grouping them together was not optimal for our use case since boolean variables and numeric variables have different intervals and properties that concern us during our analysis.
This allows us to have a clear separation of arithmetic and boolean expressions. This matches perfectly with the way SlithIR represents conditional structures, the truth value of the condition is first stored in a temporary variable. After that, the `CONDITION` SlithIR instruction is called and receives that temporary variable as its only parameter. Boolean variables that are subject to such calls have a special flag set in our representation of analysis information.

To integrate with Z3, we make use of the `exec` builtin from Python. In this way, we can dynamically execute code that adds variables and constraints to the Z3 solver. Although the adapter implemented for Z3 will not work with any other solver without adjustments, only the code that is generated and executed dynamically needs to be changed. Due to our use of abstract classes and polymorphism, the current adapter can easily be changed with the newly implemented one. The flow of constraint checking will stay the same. 

The flow of using a solver is the following:
\begin{enumerate}
    \item Query analysis information to obtain all numeric variables that are present in boolean expressions with `condition` flag set;
    \item Query analysis information to obtain all boolean expressions that are marked as conditions;
    \item Declare all numeric variables in solver;
    \item Add interval constraints for all numeric variables that do not have the default interval for their type;
    \item Add boolean expressions to the solver;
    \item Check state satisfiability;
    \item Decide if the current node is reachable or not from the current path.
\end{enumerate}
An example of output that can be further investigated by the user is presented in the following sections.

To reach this state of implementation, we faced many challenges. Besides deciding the architecture and overall solution organization, being able to translate from the Slither representation to our representation and then to a representation that is compatible with the solver was particularly challenging. This required a considerable amount of empirical study of how Slither models different types of statements, variables, and data types. Due to the fact that we want our implementation to be easily extensible, we did not model our program state with any particular solver or other integration in mind. We chose the best overall approach to model program state. This makes it compatible with any integration, but extra steps must be taken to achieve it.

Another particularly challenging aspect was correctly identifying the constraints for each program point. Accumulating constraints as we iterate through the CFG seems like a natural approach. This solution is not semantically correct, since the CFG representation of a loop contains an edge from the end point of the loop to its start point. If we only accumulated constraints as we searched the CFG, that would have led to incorrect constraint representations for programs that contain loops.

We had to define a more complex constraint flow. Constraints can only flow downward in the control-flow graph. When processing an edge that has the end of a loop as it's source edge and the start of a loop as it's destination label, constraints are not carried over. We consider this to be the most stable approach since the start point of a loop is also reached from the initial flow of a program without the loop invariant.

\section{Evaluating our solution}
\label{sec:evaluation}
Our implementation is currently able to provide a state representation for each statement of a function. This representation is written in a text file and it is presented in a human-readable format. The constraint system that we implemented is currently the strong point of our implementation when compared to other static analysis tools. We evaluated our solution on a number of functions that contain arithmetic expressions, boolean expressions, loops, conditionals, and builtin functions (\texttt{require}, \texttt{assert}, \texttt{send}, \texttt{transfer}, \texttt{call}). A contract test suite, as well as an executable build of our tool will be available at \url{https://github.com/CDU55/FROM2024VulnerableSmartContracts}.

A good example that highlights the strength of our tool is the following

\begin{lstlisting}[language=Solidity]
pragma solidity 0.8.23;

  contract DepositContract {

      mapping(address=>uint) public  deposits;
      function deposit() public payable {
          require(msg.value > 0);
          deposits[msg.sender]=deposits[msg.sender]+msg.value;
      }

      function withdraw() public  payable {
         require(deposits[msg.sender] > 0);
         payable(msg.sender).transfer(deposits[msg.sender]);
         //MISSING: set deposits[msg.sender] to 0
         //this assert statement will always revert 
         //and the users cannot withdraw ether
         assert(deposits[msg.sender] == 0);
     }
}
\end{lstlisting}

We focus mainly on the `withdraw` function. Our constraint-based approach can correctly identify the fact that the endpoint of the function is not reachable. This is due to two contradictory constraints. Having both `deposits[msg.sender] > 0` as a constraint at the start of the contract means that the caller's balance will always be positive, a usual condition for such functions. Since `deposits[msg.sender]` is not altered during execution, the constraint is still valid. When reaching the \texttt{assert} statement, the contract is led into an invalid state. The balance of a user cannot be greater than 0 and equal to 0 at the same time. This leads to a `Locked Ether` vulnerability case. Users can deposit currency and cannot withdraw it later.

The program state computed by our tool for the statement on line 15 in the example above is displayed in Figure~\ref{fig:programState}.
\begin{figure}[h]
    \centering
    {\small
\begin{alltt}
        Node 3:EXPRESSION assert(bool)(deposits[msg.sender] == 0)
         \{
        Numeric variables:\{'block.timestamp': block.timestamp uint (1,inf)
        , 'block.difficulty': block.difficulty uint (1,inf)
        , 'block.number': block.number uint (1,inf)
        , 'msg.sender': msg.sender uint (1,inf)
        , 'msg.value': msg.value uint (1,inf)
        , 'deposits': deposits:
        \},
        Booleans variables:\{'TMP_3': TMP_3 bool deposits[msg.sender] > 0 assert/require
        \},
        Constraints:[('TMP_3', True)]
        \}
\end{alltt}
}
    \caption{Example of state output for a program node}
    \label{fig:programState}
\end{figure}

The summary provided by our tool highlights the fact that the constraint imposed by the assert statement contradicts an already existing constraint. This becomes clear when reviewing the code of the \texttt{withdraw} function along with the approximated program state.

Currently, our solution is not able to automatically signal issues, the only defect that it is able to automatically find is `Unreachable code`. Thus, it can only be used as a helper tool for smart contract auditors at the moment. Automatic fault detection will be implemented in the future. The methodology that we implemented for evaluation was to run the tool, obtain the program state, and check if the targeted issue could easily be identified by the user when checking the output file. 

\textbf{Issues that could be identified by checking the output of our program are the following}:
\begin{enumerate}
    \item Array out of bounds
    \item Division by zero
    \item Unreachable code (Tautologies and contradictions in conditional instructions)
    \item Missing validation for parameters and state variables
\end{enumerate}

Table~\ref{tab:solutionEvaluation} displays how our tool performed compared to Slither on our contract collection. The contracts range from one-line functions that only contain the issue to more complex scenarios. If the issue can be reasonably deduced from the program state computed by our solution, then we consider it able to help in detecting that issue. We do not make any claim that our collection is an exhaustive one. However, these contracts highlight scenarios where Slither does not perform ideally.

\begin{table}[h]
    \centering
    \begin{tabular}{|c|c|c|c|c|}
        \hline
        Our solution & Slither & Contract & Function & Issue \\
        \hline
        \cmark  & \xmark     & BidContract  & bid  & Unreachable code  \\
        \hline
        \cmark  & \xmark     & DepositContract & withdraw & Locked Ether \\
        \hline
        \cmark & \xmark   & DivideByZeroMinimal & divide & Possible division by 0 \\
        \hline
        \cmark & \xmark   & ImproperDataValidation & participate & No parameter validation \\
        \hline
        \cmark & \xmark   & OutOfBoundsArrayMinimal & getArrayElement & Array out of bounds \\
        \hline
        \cmark & \xmark   & DivisionByZeroArray & getSomeResult & Possible division by 0 \\
        \hline
    \end{tabular}
    \caption{The results of our evaluation}
    \label{tab:solutionEvaluation}
\end{table}
\section{Current limitations}
\label{sec:limitations}
Even though we made significant progress towards a working solution, it still has limitations and we already theorized ways in which they could be mitigated. We have identified two categories of limitations: current implementation limitations and interval analysis limitations.

The first major limitation that our tool currently has is tied to the complexity of the contracts that it can process. Even though Solidity is smaller in the size of its instruction set than other well-known programming languages such as C, it still has plenty of features that must be interpreted. Since we implemented our own processors for the semantics of each instruction type, there is still work to be done. The common instructions among most programming languages are currently implemented, as well as the most used builtin functions and variables that are specific to Solidity. There is still work to do regarding other language-specific features such as `\texttt{abi.encode()}` and `\texttt{readInt8()}`. This problem will be mitigated by adding semantics for elements of Solidity that are currently missing.

Another feature that is currently missing from our tool is interprocedural analysis. Out implementation that does not yet support calls between functions that are not Solidity builtins, meaning that it is limited to intraprocedural analysis. This issue will be mitigated by implementing a recursive approach to our current analysis algorithm.

We must also mention the limitations that interval analysis imposes. This is mostly related to the precision of approximation for the interval of each variable. If the intervals are not precise enough, an issue could only be signaled as "potentially present", not as "certainly present". Increasing precision by altering the $\top$ and $\bot$ values, which are used as the default values for variables could make the implementation more prone to mistakes. This issue could be mitigated by receiving more precise default intervals from the user via annotations or a configuration file. Another way that this issue could be mitigated is by adding more types of static analysis to help the already existing one. 

\section{Conclusions}
\label{sec:conclusion}
In this progress report, we present our latest developments towards a new analysis tool that aims to find issues that are not currently detected by state-of-the-art tools. We provide a detailed presentation of how Slither runs and a set of instructions regarding how to build a new solution by leveraging the modules that it provides. We describe in detail how our tool is designed and how it interacts with third-party systems. Our implementation of interval analysis is the solution's strong point currently, but it can easily be extended to other methods of static analysis and other third-party components. At the moment of writing, our tool can provide a summary of a program's states. This summary allows the user to easily identify issues currently not detected by other similar tools, two such defects are described in this paper. We also highlight our current limitations and present our plans to overcome them.

\section{Bibliography}

\nocite{*}
\bibliographystyle{eptcs}
\bibliography{bibliography}
\end{document}